RESEARCH PROJECT

# Application of Failure Modes and Effects Analysis in the Engineering Design Process

by

Hengameh Fakhravar

ENGINEERING MANAGEMENT AND SYSTEMS ENGINEERING

OLD DOMINION UNIVERSITY
November 2020

ABSTRACT

Failure modes and effects analysis (FMEA) is one of the most practical design tools implemented in the product design to analyze the possible failures and to improve the design. The use of FMEA is diversified, and different approaches are proposed by various organizations and researchers from one application to another. The question is how to use the features of FMEA along with the design process. This research focuses on different types of FMEA in the design process, which is considered as the mapping between customer requirements, design components, and product functions. These three elements of design are the foundation of the integration model proposed in this research.

The objective of this research is to understand an integrated approach of FMEA in the design process. Significantly, an integration framework is developed to integrate the design process and FMEA. Then, a step-by-step FMEA-facilitated design process is proposed to apply FMEA along with the design process.

# TABLE OF CONTENTS

# 1.    CHAPTER 1 - INTRODUCTION

Failure modes and effects analysis (FMEA) is a tool to analyze and control the potential impacts of failures (Anleitner, 2010). Through the use of FMEA, it is expected to predict the possible failures from a product or system before it actually occurs. In this way, engineers can improve the design by deliberately controlling the causes of failures or decreasing their negative effects. One key benefit to use FMEA in product design is the "Factor of 10 rule" that early design improvements can minimize the expensive cost of modifications at the later stages of product development (Carlson, 2012, pp. 5-6).

The origin of FMEA rise from the industrial initiative, particularly from automotive industries and the military. The procedure of FMEA can be efficient in a particular industrial sector. Yet, significant adapting efforts are required if we want to use the same process from one application to another application. So, different standards related to FMEA have been proposed by some organizations such as the military standard of the United States (MIL-P-1629), Society of Automotive Engineers (SAE: ARP5580).

On the other hand, academic researchers have devoted significant efforts in the research of engineering design in the past few decades. The final report can be found in some notable articles by Pahl et al. (2007), Dieter and Schmidt (2009), and Ulrich and Eppinger (2012). Intuitively, a design might be primarily referred to as the artifact (physical structure or software) that is perceived by the users directly. Yet, this article on engineering design has emphasized the significance of functions to explain the artifact as part of the design process. That is, the design information is fundamental before a design concept is developed.

In this research project, the process of design is considered the mapping process between design elements. Three types of design elements are mainly considered in this research: product functions, customer requirements, and design components. Product functions are referred to as the intents of the design without stating the specific solutions. For instance, if we want to eat the can's food, the function can be explained as "open the can." There can be different ways to open the can.

Customer requirements are considered as the input data snd information indicating the needs of customers, and they are usually gathered and organized in a marketing department through the market research activity. Design components suggest that the specific design solutions that are implemented to achieve the product functions. (Tahami & Fakhravar, 2020b, 2020a)

In brief, this research project considers the design process as mapping "requirements to functions to components." This mapping references the methodology of Quality Function Deployment (QFD) in quality management and Axiomatic Design (AD) in engineering design. Mostly, QFD can involve mapping matrices from customer needs to engineering characteristics and then to part characteristics (Hauser and Clausing, 1988). In AD, the mapping framework has been proposed for four design areas: customer, functional, physical, and production (Suh, 1990).

By considering FMEA as a design tool, the research question is how to systematically employ the features of FMEA in a design process. The academic research in engineering design has provided a solid foundation for critical design concepts such as requirements, functions, and components. This research project is to systematically interpret these design concepts in the context of FMEA. In such a way, the use of FMEA can be carried along with the design process. Expectedly, the information on failure analysis (from FMEA) can be employ at the later stages of the design process. This practice can improve the consistency of the failure information among different teams in product development.

The research approach has two aspects. The first one is related to the design process. It is required to clearly define design elements and how these design elements are related to each other. This provides the base for the integration effort with FMEA. The second aspect is related to FMEA. The key is to specify the definition of the contents of failure modes concerning design elements. The failure modes are specified in a design context; the procedure of FMEA could be adapted in the design process for analyzing the risks.

The goal of this research is to research an integrated approach of FMEA in the design process.

Two objectives are specified for this research project.

Objective 1: Explain a framework so that both FMEA and the design process can be integrated in a consistent manner.

Objective 2: Explain the FMEA-facilitated design process that consists of a step-by-step approach to applying FMEA along the design process.

# 2   CHAPTER 2 - LITERATURE REVIEW

## 2.1 DEVELOPMENT OF FMEA IN INDUSTRY

The foundation of FMEA can be traced back to the military standard of the United States in 1949 (MIL-P-1629), and the process at that time was titled "Failure Mode, Effects and Criticality Analysis" (FMECA). This military standard was revised in 1980 (Military, 1980). According to the forward of this standard, the motivation of using FMECA is to recognize the failure modes for assessing the "adequacy and feasibility" of the design and supporting the design decision. Beyond the military applications, FMECA has been applied for space missions. One example is that the National Aeronautics and Space Administration (NASA) has used FMECA for the Apollo program to examine the "crew safety problems and system unreliability" (NASA, 1966). The difference between FMEA and FMECA is that FMECA additionally consider a criticality analysis. For each failure mode, criticality analysis considers the possible failure, mode ratio of unreliability, and the probability of loss (Tahami et al., 2019 Carlson, 2012,).

The large scale use of FMEA can be found in the automotive industry. The Society of Automotive Engineers has introduced the FMEA standard in 2001 (SAE 2001). As noted in (Carlson, 2012) and (Bertsche, 2008), the Ford Motor Company has been the first automobile.

Manufacturers in the 1970s tried to apply FMEA and FMEA as one of the important tools in reliability analysis. Also, FMEA has been recognized as a critical tool in quality engineering as it is part of the body of knowledge to the Black Belt Certification (ASQ, 2013; Creveling et al., ,2003). International Organization for Standardization (ISO) has managed relevant techniques in one standard, ISO/TS 16949 (Duckworth and Moore, 2010, p. 51). The paper indicates that the evidence about the usefulness and popularity of FMEA in industrial practice. Beyond the mechanical systems and aerospace, other industrial sectors have reported the application of FMEA for reliability and quality analysis, such as software (Reifer 1979) and healthcare (De Rosier et al., 2002)

According to military standards, the techniques of FMEA initially targets the improvement of design. By identifying the possible failure modes during the design stage, we can minimize the impacts of these failure modes by improving the original design. This practice can reduce the total cost as compared to the later stages of modifications at the end of the product development process. This area of FMEA has been termed as "Design FMEA" (Anleitner, 2010). When FMEA becomes familiar in industrial practice, engineers have adapted and extended the FMEA techniques to other areas. Three examples of different FMEA fields are listed and briefly explained as follows.

1- Process FMEA (Teng and Ho, 1996): the Process FMEA is used to analyze the risks related to the manufacturing operations.

2- Concept of FMEA (Carlson, 2012): the Concept FMEA is used to assess the reliability and safety of the design concepts during the multi-criteria selection process.

3- Social Responsibility FMEA (Duckworth and Moore, 2010): Social Responsibility FMEA is used to evaluate a company's operations in the aspect of social responsibility. In that way, failure modes are referred to as minimum impacts on the society and environment as a whole. It is supposed to enhance social responsibility performance so that the company can explore alternative ways for the solutions for a better environment and society. (Tahami et al., 2016)

After using FMEA in industrial practice, some researchers propose a systematic FMEA methodology to build fundamental principles. Also (Anleitner, 2010) has proposed the deductive design FMEA method that specifies the inputs and outputs of each procedural step. (Carlson, 2012, p. 18) has also designed the relationship diagram between Design FMEA and Process FMEA by systematically connecting the required input and output information.

Based on this review, two observations are made about FMEA in industrial practice. Firstly, originated from the design field, the application of FMEA has been extended to other areas in product development. From a manufacturer's viewpoint, risks are not isolated. Such an extension provides an opportunity to integrate the information of risks across different domains.

Secondly, first FMEA procedures try to provide practical but somewhat ad-hoc guidelines. Thus, the actual outcomes from FMEA can significantly from one project to another. Recent research has focused on formalizing the FMEA procedure to promote the utility of FMEA in the actual practice. This research project corresponds to these two observations by integrating the requirements, components, and functions in an FMEA framework.

## 2.2. FMEA IN ACADEMIC RESEARCH

In academic research, one of the research directions related to FMEA is to improve the features of FMEA for more comprehensive failure and risk analysis in product development. (Stone et al., 2005) suggested the function-failure design method (FFDM) to support FMEA in the conceptual design stage. The original technique behind FFDM is to use the notion of functions to systematically define failure modes in engineering design. (Chao and Ishii, 2007) presented the error-proofing method by adapting FMEA to prevent design errors, which are classified into six categories: knowledge, communication, analysis, execution, change, and organization. FMEA is applied to guide engineers to explore the possible errors in these categories through the question-asking techniques. (Tahami & Fakhravar, 2020c)

One specific issue of research about FMEA is to tackle the accuracy and appropriateness of the risk priority number (RPN). In brief, RPN is the product (i.e., by multiplication) of three numerical rankings: severity of risks, the occurrence of risks, and control of risks. The higher ranking values show the worse risk situations, and RPN is used to prioritize the failure modes due to their risk situations. Then, the issue of RPN is that the same value of RPN can represent different risk situations. For instance, if the RPN value is equal to 600, we actually could not decide whether the risk situation is (1) low occurrence and high severity or (2) high occurrence and low severity. It is because numerical rankings are simply aggregated into one value (i.e., RPN).

To address this specific issue of FMEA, some researchers have presented more comprehensive approaches to compute RPN and prioritize failure modes. (Pillay and Wang, 2003) applied fuzzy sets and rules to infer and prioritize the risk situations, and their approach allowed users to explain more specific scenarios of risks for particular contexts. (Kmenta and Ishii, 2004) used the probability and cost as a common basis to evaluate mate and prioritize the risk situations in a more precise manner. (Chang and Cheng, 2011) used the fuzzy concept ordered weighted averaging (OWA) that allowed weighting factors and human imprecision in the assessment of risk situations. (Bradley and Guerrero, 2011) presented a data-elicitation technique integrated with the interpolation algorithm to support the risk assessment in FMEA. (Chang et al., 2013) presented an exponential risk priority number (ERPN) for providing unique numerical values mapped to different risk situations.

This research project will not address the issue of RPN. Instead, this research project focuses on the methodology for integrating FMEA in the design process.

### 2.3. THE GAP BETWEEN ACADEMIC AND  INDUSTRIAL FMEA

From the discussion of FMEA in industrial experience, it is stated that the application of FMEA becomes more popular in the industry. Because FMEA often requires the effort of a team, the FMEA procedure needs to be more accurate and systematic for better communication and ensuring quality outcomes. Recent articles, such as (Anleitner, 2010) and (Carlson, 2012) have supported the direction of FMEA development in the industry. Additionally, it mentions that FMEA has been extended in different stages of product development for the risk analysis (e.g., process FMEA and requirement FMEA). Such an extension effort remains active in assisting the organization of the process of product development.

From the discussion of FMEA in academics, while this research project does not provide new approaches for handling RPN, it expands the research works of (Stone et al., 2005) and (Chao and Ishii, 2007) by combining more design elements in the practice of FMEA. Particularly, the design process of this research project shows a flow from customer requirements to design functions and components (i.e., requirements to functions to components). Essentially, this kind of design flow is like the methodology of Quality Function Deployment (Suh, 1990) or Axiomatic Design (Hauser and Clausing, 1988).

# 3 CHAPTER 3- INTEGRATION FRAMEWORK

This chapter aims to explain the integration model that provides a platform for joining the information from the design and FMEA areas. The information related to the design and FMEA area is first separately discussed. The integration of these areas is based on the definition of failure modes based on the design elements.

## 3.1. BASIC ELEMENTS IN ENGINEERING DESIGN

The design process in engineering has not yet been standardized among practitioners and researchers, and thus many terms and design processes can be found from literature (Carlson 2012; Creveling et al. 2003; Pahl et al. 2007). In this research project, three representative and foundational elements in engineering design are considered: requirements, functions, and components.

Requirements are referred to the ultimate needs that are used to inspect the product's goodness. In quality engineering, requirements are mostly referred to customer needs, where the concept of customers can involve multiple aspects, and the following list provides some examples.

1- The customers who purchase the product
2- The clients who use the product
3- The governmental policies and regulations
4- The environmental requirements

Generally, requirements can be viewed as the external and somewhat non-technical expectations that are required for the products. Product failures can be claimed if the requirements of a product cannot be met. This highlights the significance of requirements in engineering design.

The concept of functions in engineering design might come from the philosophy "Form Follows Functions" in architecture (Pahl et al., 2007). The idea of functions is to distinguish between "what" and "how" in design. Particularly, "what" describes the basic aim (or functions) of the design, e.g., separating a piece of paper into a two-piece. Then, "how" describes the solutions to achieve these purposes, i.e., use a scissor as one way to separate a piece of paper. The significance of the function is that multiple solutions are feasible to achieve the same function. For instance, instead of using a scissor, we can use a ruler or our hands to separate the piece of paper. At this point, the concept of functions can help engineers to think of various solutions without committing to any solutions too quickly creatively.

(Stone et al., 2000) is used the function concept based on a functional basis for the design. Particularly, a function is expressed in a phrase structure "verb + noun" to highlight the "action" to be carried in a function. Then, each action can be viewed as a transfer function that converts some input into Some output. For instance, "separate a piece of paper into two" is a function with an input "one piece of paper" and an output "two pieces of paper." The inputs and outputs are further classified into three types: material, energy, and information. Also, functions could be decomposed by explaining how some sub-functions are required to achieve a high-level function. At the point, a product could be described as a set of sub-functions that are connected to deliver its major functions. To manage the sub-functions, a functional block diagram is commonly used in practice. Further details of functional design and analysis could be found (Kossiakoff et al., 2011).

Components are referred to the solutions that are chosen to satisfy specific function(s) or sub-function(s). A component is basically a concept that is presented by (Ulrich and Eppinger, 2012, p. 98) as "an approximate description of the technology, form of the product, and working principles." For instance, suppose that a scissor chooses as the "component" to separate a piece of paper. Then, the scissor concept roughly implies the use of two blades for cutting, and further engineering details are necessary for further implementation (e.g., size of scissor, blade materials, etc.) Yet, the choice of a concept confines the direction of engineering efforts, and it is often considered as a critical decision towards the success of a design (Ulrich and Eppinger, 2012).

After defining the concept of requirements, functions, and components, this research focuses on the organization of these three types of elements along with the product development process. The first step of the organization is to itemize these elements exactly as the design information of a product. Let R, F, and C be the set of requirements, functions, and components, respectively, and these items are denoted as follows.

1- Requirements: R = {r1, r2, …, rm}, m is the total number of requirements

2- Functions: F = {f1, f2, …, fn}, n is the total number of functions

3- Components: C = {c1, c2, …, cp}, p is the total number of components

Additionally, to the design information, FMEA targets the failure information of a product.

## 3.2. BASIC ELEMENTS OF FMEA

Though FMEA is originated from a military standard (Military, 1980), the practice of FMEA becomes diversified by incorporating different documentation styles and ranking schemes (Carlson, 2012). However, there are some basic elements that characterize the fundamental principles of FMEA. In this research, these basic elements are causes, effects, and failure modes.

Based on (Carlson, 2012, p. 28), a failure mode can be defined as "the manner in which the product or operation failure to meet the requirements." In this sense, a failure mode should be a simple description of the failure without stating the cause behind it or the effect after it. In the engineering design context, a failure mode could be referred to any dissatisfaction related to requirements, functions, and components. Towards the integration efforts, one main idea of this research is that a failure mode should be defined based on the known requirements, functions, and components of the design. This idea provides guidance for engineers to prepare logically the FMEA documents related to the design context.

The casual analysis in the study of engineering failures is a challenging step, and fault tree analysis (FTA) is usually identified as one common tool for this kind of task (O'Connor, 2012). Compared to FTA, FMEA does not focus on the casual relationships of failures. But, FMEA suggests identifying the causes and effects associated with every failure mode. Mainly, the causes are the rational reasons that can lead to the occurring of the failure mode, and the effects represent the negative impacts if the failure mode is materialized. In this case, the causes and effects are connected by failure modes only without involving the chain effects between causes and effects.

For instance, when engineers make a failure analysis of a phenomenon, "a user cannot take a photo," the analysis process in FTA may be similar to Figure 1, a top-18 down analysis method to find the answers (i.e., root causes). In FTA, all events under the phenomenon are viewed as possible causal factors. Meanwhile, one event can result from another event at the lower level or a reason for the other event in the upper level. Therefore the causal relationships of failures are broke down. One causal factor is possibly rooted out in different branches in FTA.

From Figure 1, we can find that "camera module is damaged" is the upper event of the causal factor "lack of R/C components for protection", and it is a causal factor for the event "camera module cannot be executed." Also, "camera module is damaged" can be a root out under the analysis branch of "Design factors" or under the branch of "Manufacture factors."

However, the cause-failure mode-effect chain is clearer in FMEA. In a similar case, the failure mode is identified as "Camera module is damaged" if the analyzed items in FMEA is components, the possible causes is "lack of R/C components for protection" and "Incorrect circuit design"; the possible effect is "Camera module cannot be executed." Similarly, if the analyzed item in the FMEA is functions, the failure mode is "Camera module cannot be executed", and the causes are identified as "Camera module is damaged", "Lack of power supply", "Incorrect patter", or so on; the effect is identified as "A user cannot take photos". Instead of considering "what if" continuously to build up the effect results or to ask "why" continuously to find out the root causes, engineers focus on the direct effects and the causes of the failure modes corresponding to the analyzed items in FMEA.

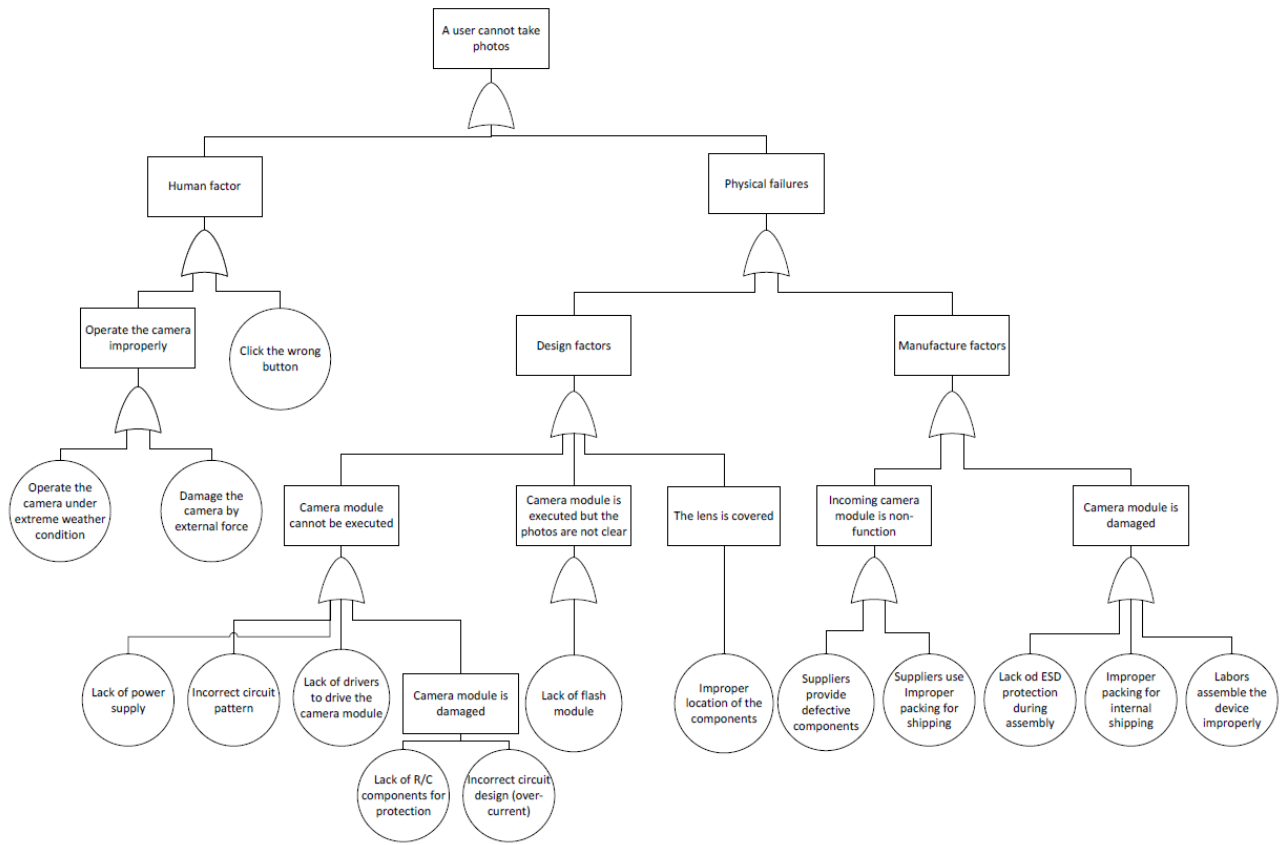

Figure 2: Fault tree analysis (FTA) diagram example

After explaining the implication of failure modes, causes, and effects, this study focused on the design of these three types of factors that are relevant to the product development process. The first step of the organization is to itemize elements explicitly as the failure information of a design.

Let FM, CA, and EF be the set of failure modes, causes, and effects, and these items are further denoted as follows.

1- Failure modes: FM = {$fm_1$, $fm_2$, …, $fm_q$}, q is the total number of failure modes

2- Causes: CA = {$ca_{ij}$}, $ca_{ij}$ is the jth cause of the ith failure mode

3- Effects: EF = {$ef_{ik}$}, $ef_{ik}$ is the kth effect of the ith failure mode

In FMEA, the understanding of failure modes is the major step to investigate the corresponding causes and effects.

### 3.3. DEFINITION OF FAILURE MODES FROM DESIGN ELEMENTS

The research effort of this concept is about integrating the design process with the utilize of FMEA. The heart of this integration lies in the definition of failure modes based on the design information. From Section 3.1, three design elements (i.e., requirements, functions, and components) are subdivided to form the basis of knowledge. Simultaneously, a failure mode could be viewed as "the way" in which the product fails. The basic goal is by classifying their typical "failure manners," we can define the failure mode for each type of design element.

When product requirements are considered customer needs, the corresponding failure modes can generally be viewed as failing to meet the requirements. More specifically, there could be five different manners (or modes) in which failures could take place related to requirements.

These modes are listed and explained as follows.

- ❖ Absence: the requirement is totally not met

- ❖ Incompleteness: the requirement is only met partially

- ❖ Intermittence: the requirement cannot be smoothly met

- ❖ Incorrectness: the requirement is met incorrectly

- ❖ Improper occurrence: the requirement is met at the wrong time

For instance, consider that one requirement of the smartphone example is "have an internet connection at all times." Then, the engineers can investigate which manners this requirement can fail, and three potential failure modes can be identified as follows.

- ❖ Absence: the smartphone cannot support an internet connection

- ❖ Intermittence: users experience frequent interruptions with an internet connection

- ❖ Improper occurrence: it takes a long time to connect to the internet

Note that engineers only investigate the failure modes that are logical in the context of requirements. We can skip some modes that may not make sense toa particular need. For instance, it is found that "incompleteness" is not very reasonable in relation to "internet connection," and thus, it is not considered as one failure mode. So the value of the five modes is to provide the guideline for engineers to carry FMEA systematically.

Similarly, functions are expressed in the structure of the phrase "Verb + Noun," and thus, the methods of failure of tasks are considered negative phrases when compared to the active verbs used to express a function. Moreover, while we emphasize the verbs in the function description, the failure modes are considered as any negation from its original description. In this research project, we categorized the failure modes of functions into five types.

❖ Malfunction: the function is not executed

❖ Interference: the function execution has interfered

❖ Decayed: declined function performance; the function execution doesn't reach the standard after a certain time

❖ Incompleteness: the function is partly executed

❖ Incorrectness: the function is incorrectly executed

For instance, if the function "display images" is being studied, two failure modes are potentially identified by engineers.

❖ Malfunction: the smartphone does not display images

❖ Interference: the image display has interfered

The failure modes of the component in this research project are specified for electronic components. The electronic components are really sensitive to the design criteria, and any details of the design can have effects on the components' performance. Yet, only the most serious failure modes which stop components from performing the designed functions are considered in this research project, and we categorized them as below.

❖ Damaged: components lose abilities to achieve the functions

❖ Loss of efficiency: the components perform functions less efficiently than its technical specifications

❖ EMI: the components emit radiation

❖ Non-compatible: a component's specification is non-compatible to perform the function properly

For instance, a GSM transceiver is studied, and the engineers possibly identify three failure modes listed below.

❖ Damaged: the GSM transceiver is damaged (might be burned out or discharged)

❖ Loss of efficiency: the GSM transceiver has difficulty to access the GSM network even when it is powered

❖ EMI: the GSM transceiver emits the radiation

The symbols are used to denote the failure modes for each type of design element, respectively.

❖ $FM_a$ = the $a$th failure mode (e.g., $FM_1$)

❖ $FM_{ari}$ = the $a$th failure mode of the $i$th requirement (e.g., $FM_{1r1}$)

❖ $FM_{afj}$ = the $a$th failure mode of the $j$th function (e.g., $FM_{1f1}$)

❖ $FM_{ack}$ = the $a$th failure mode of the $k$th component (e.g., $FM_{1c1}$)

# 4 CHAPTER4- FACILITATED DESIGN PROCESS

The purpose of this chapter is to propose a process of facilitating the implementation of FMEA in the design process. In the beginning, the three-phase design process is provided based on the three elements in engineering design (i.e., requirements, functions, and components). In accordance with this design process, a process is developed that incorporates the FMEA practice.

## 4.1. THREE-PHASE DESIGN PROCESS

The three-phase design process focuses on the acquisition of the design information as part of the design efforts. In Particular, the design information in this research is referred to as the design requirements, functions, and components. Significantly, the development of this design information is similar to the process of mapping from the customer domain to the functional domain and then to the physical domain in axiomatic design (Suh, 2001). The research work emphasizes how to acquire the information of requirements, functions, and components systematically. The three phases of the process are labeled and listed as follows.

- ❖ Phase 1: Identification of requirements

- ❖ Phase 2: Deployment of product functions

- ❖ Phase 3: Definition of product components

### 4.1.1. IDENTIFICATION OF REQUIREMENTS

In Phase 1, the major task is to identify the list of customer requirements that can characterize the way of product development. (Evans and Lindsay, 2005) presented that the importance of requirements has been studied in six-sigma management and quality engineering. Currently, market research has traditionally played an important role in collecting and analyzing expectations, needs, insights, and customer preferences. Typical strategies from market research include formal surveys, focus groups, and online monitoring (Bradley, 2010).

In addition, understanding from designers and engineers is also important, depending on their knowledge and experience. Meeting customer expectations is often considered the minimum required to achieve a customer satisfaction line. Competitively, it is necessary to please customers beyond the expectation, so the developer's understanding of industry trends and challenges is very important in identifying the requirements.

In particular, the understanding of engineers is important in determining the exciters/delighters in the Kano classification system (Evans and Lindsay, 2005). In addition, the insights of engineers ensure that the results of market research can be translated into technical aspects without deviation.

Furthermore, the information on existing products could assist the identification of requirements. By studying the existing products from the competitors in the same industry, we can compare the technical performance, product features, and other characteristics. This technique is called competitive benchmarking that is common for setting competitive and realistic goals in product development (Stapenhurst, 2009).

The expected outcome of phase 1 is a list of requirements that characterize the directions of product development. To get this point, the method of quality function decomposition (QFD) matrix is used. QFD is a method that focused on listening carefully to the voice of the customer and then responding effectively to those needs and expectations (Evans and Lindsay, 2005).

Some different types of matrices are designed for different purposes of practicing QFD, such as product characteristic matrix, requirements matrix, the design matrix, and so on. For the purpose of identifying the requirements here, the requirements matrix is used to transform the customer requirements to design requirements, shown in Figure 2.

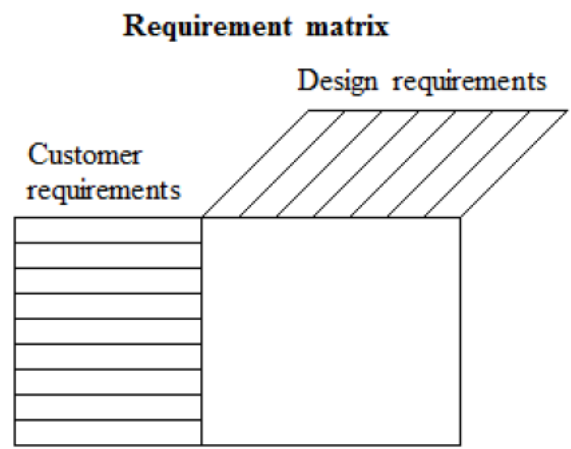

Figure 3: Requirement matrix

### 4.1.2. DEPLOYMENT OF PRODUCT FUNCTION

Depending on the requirements, product functions are deployed to indicate the engineering goals that are expected in the product. Delivery of product functions can be seen as translating product descriptions from customer languages into engineering languages. In the development of the product, this process is similar to the QFD design matrix, which involves the mapping between requirements and engineering features (ASQ, 2004). However, it should be noted that engineering features in QFD are referred to as some measurable quantities that are relevant to the satisfaction of customer requirements.

To deploy the functions based on requirements, we can use some question-asking techniques that are not uncommon in conceptual design. Listed below are two possible questions that may useful to deploy functions from requirements.

❖ What are the expected inputs and outputs if the requirement is satisfied? The process between inputs and outputs can be defined as the function(s).

❖ What product do actions need to be carried out in order to satisfy the requirement? Actions here can be expressed as the "verbs" of the functional descriptions.

After deploying the product functions, two specific outputs are expected: a set of functions and the mapping between requirements and functions.

Referring to Section 3.1, the set of functions is denoted as $F = \{f_1, f_2, \ldots, f_n\}$. The mapping could be expressed in a matrix denoted as $RF = \{rf_{ij}\}$, where $rf_{ij}$ is equal to one if the $j$th function is should satisfy the $i$th requirement. Otherwise, $rf_{ij}$ is equal to zero.

### 4.1.3. DEFINITION OF PRODUCT COMPONENTS

Depending on the list of functions, product components are defined to specify the particular solutions to achieve the functions. The concept of design is often obtained after defining the components that address all the product functions. After that, the definition of product components can be viewed as the process of conceptualization in engineering design. Various conceptual production techniques have been proposed, such as morphological chart, brainstorming, and axiomatic design (Suh, 2001).

In this study, the information of functions is one important input to influence the definition of components. In particular, the inputs and outputs of each function help engineers imagine what engineering components are capable of achieving such modification. (Ulrich and Eppinger, 2012, chapter 7) have introduced various search techniques for concept generation. In addition, they have suggested some methods for obtaining design ideas in thinking, such as "make analogies" and "wish and wonder."

Defining the components to achieve the functions can be viewed as a creative design process. The adjective "creative" implies that we do not have an automated way to always lead to successful designs. Among some design best practices (Stapenhurst, 2009), engineers were suggested to focus on the functional descriptions in view of the necessary inputs and outputs. This helps them determine what exactly is to be achieved in products.

After clarifying defining the components, two specific outputs are expected: a set of components and the mapping between functions and components. Referring to Section 3.1, the set of components is denoted as $C = \{c_1, c_2, \ldots, c_p\}$. The mapping can be expressed in a matrix denoted as $FC = \{fc_{jk}\}$, where $fc_{jk}$ is equal to one of the $k$th components required to achieve the $j$th function. Otherwise, $fc_{jk}$ is equal to zero.

Significantly, the purpose of this section is not to suggest a new design process. Designs information of requirements, functions, and components are discussed extensively in the literature on engineering design and development of product (Pahl et al., 2007). In this view, the purpose of this section is to discuss in detail how we obtain the design information in this research and to explain the implications required for the integration of the approach with the FMEA.

## 4.2. FMEA EVALUATION SCHEMES IN THE DESIGN PROCESS

Risk Priority Number (RPN) is a numerical ranking of the risk on each possible failure mode, made up of the arithmetic product of the three elements: severity of the effect, the likelihood of occurrence of the cause, and the likelihood of detection of the cause. (Carlson, 2012, chapter 3). The severity is given associated with the most serious effect based on the scheme.

The detection is given based on the control plan, and it associates with the chance a cause may be detected by the control method. The schemes are specific to the companies, projects, or products (e.g., the scaling table may vary depending on the usage.) In this section, the evaluation scheme of these three numerical rankings is provided in Table 1, Table 2, and Table 3. To give the assessing number to each element, engineers followed by the evaluation scheme tables in general. In the next section, a reasoning method is developed to obtain these evaluated ranking systematically.

Definition of symbols:

❖ S = severity ranking

❖ O = occurrence ranking

❖ D = detection ranking

❖ RPN = (S x O x D)

**Table 1:** FMEA evaluation scheme for occurrence

| Rank | Occurrence |
|------|------------|
| 9-10 | Frequency $\geq 1$ in 20 |
| 7-8 | Frequency $\geq 1$ in 125 |
| 5-6 | Frequency $\geq 1$ in 1250 |
| 2-4 | 1 in 100000 $\leq$ Frequency $\leq$ 1 in 10000 |
| 1 | Frequency $\leq$ 1 in 1000000 |

(ISO MIL-STD-105E & average sales quantity per model per region)

**Table 2:** FMEA evaluation scheme for severity

| Rank | Detection |
|------|-----------|
| 9-10 | No apparent method to detect |
| 7-8 | Controlled by design analysis |
| 5-6 | Controlled by following standard design documents |
| 2-4 | Controlled by pass/fail or reliability test |
| 1 | Controlled by real-life product test (function-simulated test) |

**Table 3:** Table 2: FMEA evaluation scheme for severity

| Rank | Severity – Requirement | Severity – Function | Severity – Component |
|------|------------------------|---------------------|----------------------|
| 9-10 | Safety issue | Safety issue | Safety issue |
| 7-8 | The users choose competitors' products | The users meet difficulty to operate the function | Effects on primary function |
| 5-6 | The users need to return the device to fix the problem | The users can operate functions but the performance is under standard | Effects only on secondary function |
| 2-4 | The users might tolerate the problem and continue to use the product | Isolated defect and doesn't affect function execution | Non-functional effects |
| 1 | Invisible to a user (make no difference to a user) | Invisible to users | Invisible to a user |

## 4.3. RESONIN OF CAUSES AND EFFECTS IN FMEA

Definitions of symbols:

❖ $S(FM_{ari})$ = severity ranking of the $a$th failure mode of the $i$th requirement

❖ $S(FM_{afj})$ = severity ranking of the $a$th failure mode of the $j$th function

❖ $S(FM_{ack})$ = severity ranking of the $a$th failure mode of the $k$th component

❖ $O(FM_{ack})$ = occurrence ranking of the $a$th failure mode of the $k$th component

❖ $S(r_i)$ = severity ranking of the $i$th requirement

❖ $S(f_j)$ = severity ranking of the $j$th function

❖ $S(c_k)$ = severity ranking of the $k$th component

❖ $O(r_i)$ = occurrence ranking of the $i$th requirement

❖ $O(f_j)$ = occurrence ranking of the $j$th function

❖ $O(c_k)$ = occurrence ranking of the $k$th component

❖ $D(c_k)$ = detection and control ranking of the $k$th component

Forward analysis of effects

The reasoning of effects showed below is to obtain the severity of effects of an element (i.e. requirements, functions, and components). The purpose is to prioritize the risk consequence of the element.

❖ $S(r_i)$ = Maximum of $S(FM_{ari})$ (e.g. $S(r_1)$= Max[$S(FM_{ar1})$]

❖ $S(f_j)$ = Maximum of $S(FM_{afj})$ (e.g. $S(f_1)$= Max[$S(FM_{af1})$]

❖ $S(c_k)$ = Maximum of $S(FM_{ack})$ (e.g. $S(c_1)$= Max[$S(FM_{ac1})$]

The reasoning of effects showed below is to obtain the severity of functions (and components) from requirements (and functions.)

❖ $S(f_j)$ = Maximum of [$S(r_i)$ of all $rf_{ij}$] (e.g. if $f_1$ is deployed to satisfy both $r_1$ and $r_2$, then $S(f_1)$ is the maximum value of $S(r_1)$ and $S(r_2)$)

❖ $S(c_k)$ = Maximum of [$S(f_i)$ of all $fc_{jk}$] (e.g. if $c_1$ is defined to achieve both $f_1$ and $f_2$, then $S(c_1)$ is the maximum value of $S(f_1)$ and $S(f_2)$)

40

Backward analysis of causes

The reasoning of causes showed below is to obtain the occurrence of causes an element (i.e. requirements, functions, and components). The purpose is to know the chance of occurrence and complete the documentation of FMEAs to prioritize the risk of failure modes.

❖ O_components → O_functions → O_requirements

❖ $O(c_k)$ = max $O(FM_{ack})$

❖ $O(f_j)$ = Maximum of [$O(c_k)$ of all $fc_{jk}$] (e.g. if $f_1$ has to be achieved by both $c_1$ and $c_2$, then $O(f_1)$ is the maximum value of $O(c_1)$ and $O(c_2)$)

❖ $O(r_i)$ = Maximum of [$O(f_i)$ of all $rf_{ij}$] (e.g. if $r_1$ has to be satisfied by both $f_1$ and $f_2$, then $O(r_1)$ is the maximum value of $O(f_1)$ and $O(f_2)$)

## 4.4. METHODICAL PECEDURE

A methodical procedure is developed that incorporates the FMEA practice in this section. The procedure aligns with the general FMEA practice but the evaluation rating is supported by the reasoning based on Section 4.3. The procedure is showed as following and a comprehensive case study based on the procedure is done in Chapter 5.

Step 1: identify the requirements and determine their failure modes

❖ Identify the requirements (based on Section 4.1.1)

❖ Determine their failure modes (based on Section 3.3)

Step 2: analyze the effects of requirements

❖ Analyze the effects of requirements and decide severity (based on Section 4.2)

❖ Determine $S(r_i)$ from reasoning by $S(FM_{ari})$ (based on Section 4.3)

❖ Prioritize the "risk consequences" of requirements

Step 3: deploy the functions and determine their failure modes

❖ Deploy the functions (based on Section 4.1.2 plus the risk consequences of requirements from Step 2)

❖ Determine their failure modes (based on Section 3.3)

Step 4: analyze the effects of functions

❖ Analyze the effects of functions and decide severity (based on Section 4.2)

❖ Reasoning $S(f_j)$ by $S(r_i)$ (based on Section 4.3)

➢ Reasoning $S(f_j)$ by $S(r_i)$

➢ Reasoning $S(f_j)$ by $S(FM_{aff})$

➢ Determine $S(f_j)$

❖ Prioritize the "risk consequences" of functions

Step 5: define the components and complete the FMEA document in the component domain

❖ Define the components (based on Section 4.1.3) plus the information from Step 4

❖ Determine their failure modes (based on Section 3.3)

❖ Analyze effects and decide severity (based on Section 4.2)

❖ Determine $S(c_k)$ (based on Section 4.3)

➢ Reasoning $S(c_k)$ by $S(f_j)$

➢ Reasoning $S(c_k)$ by $S(FM_{ack})$

➢ Determine $S(c_k)$

❖ Causes

➢ Analyze the causes of components and decide occurrence (based on Section 4.2)

➢ Determine $L(FM_{ack})$ (occurrence ranking) $\rightarrow$ determine $L(c_k)$

❖ Define current control plan and assign detection (based on Section 4.2)

❖ Complete the FMEA of component

Step 6: complete the FMEA document in the function domain

❖ Analyze the causes of functions (based on Section 3.4)

❖ Determine $O(f_j)$ (based on Section 4.3)

❖ Complete FMEA of functions

Step 7: complete the FMEA document in the requirement domain

❖ Analyze the causes of requirements (based on Section 3.4)

❖ Determine O($r_i$) (based on Section 4.3)

❖ Complete FMEA of requirements

Figure 3 below shows where above steps are incorporated in the procedure of FMEA-facilitated design process.

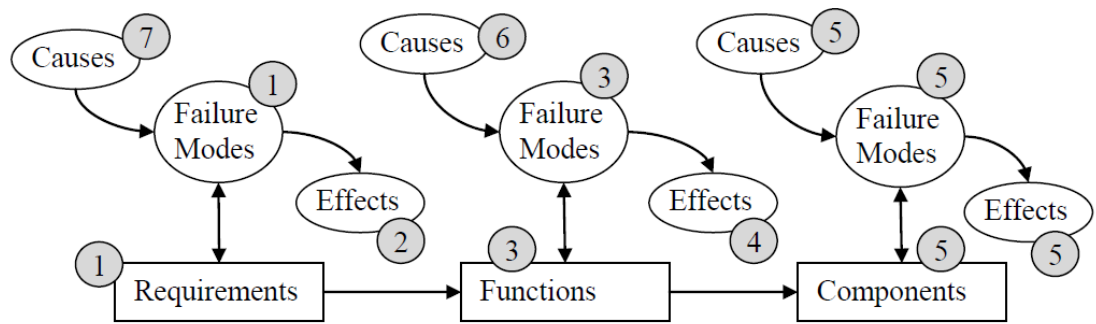

Figure 5: Schematic diagram of the procedure of FMEA- facilitated design process

After completing the FMEA-facilitated design process, benefits of the procedure can be expected in view of the design practice, listed as below.

❖ Cascade of the information of effects when performing the "component design" helps engineers prioritize not only the failure mode itself but also the risk of a component, a function, or a requirement. Besides, it motivates the engineers make decisions on "safe" solutions at some critical parts when the components selections are on the basis of the prioritization of the functions risk consequences.

❖ Logical development of FMEA documents by the support of reasoning the rating numbers minimize the mental efforts and mistakes. At the same time, the FMEA documents are completed in a relatively shorter time and more accurate results.

❖ The documentation of the FMEA of functions or the FMEA of requirements is potentially to be used for re-design or re-engineering of the product. The documents provide a guideline for engineers to assess the risk of the similar products with similar requirements or functions.

# 5.   CHAPTER 5 - CONCLUSION

For decades, FMEA is used in industry, and different approaches are proposed and improved by researchers. The practice of FMEA allows engineers to examine the design and control or eliminate the possible causes of failures. In this research project, an integration framework is developed to integrate the FMEA and engineering design process, and the goal is to utilize the failure analysis information from one stage to another stage of the design process. To reach this goal, it is necessary to integrate different FMEA documents that are implemented throughout the product development process. In this research project, the engineering design process is defined by specifying three key design elements: requirements, functions, and components. It is by the reference of QFD and AD methodologies. These design elements are treated as a foundation of the proposed integration model.

This research project aims to apply FMEA along with the design process, and thus a step-by-step FMEA-facilitated design procedure is proposed. The procedure is then demonstrated and verified by a case study. From the case study, we have three observations and also notice two limitations to the integration model.

The first observation is the information of failure analysis can be utilized at the later stages, and it ensures the failure analysis is consistent. The second observation is that the practice or prioritizing the risk consequence of requirements and functions helps engineers to select components and design the product according to the consideration of the risk consequence that the engineers listed in the earlier analysis process. The third observation is the completion of FMEA in the domain of the component promotes the documentation of the FMEA of functions and FMEA of requirements. Nevertheless, there are two limitations in applying the integration model in real design practice. The detection assessing value is required in FMEA practice most of the time. Yet, this research is unable to capture the real control method due to the lack of information. Instead, we made an assumption on the control and detection part in the FMEA of the component domain. Secondly, due to the limitation of time, component relations are not mapped in this research project. As a result, some causes of function failure modes are unable to be observed by only reviewing the failure analysis of components and the mapping between functions and components. These two limitations are left to be addressed in future research.

Notably, the contributions of this research project are pointed out below.

1- The integration of different types of FMEA documents in the product development process, which has not been found in the literature

2- A step-by-step FMEA-facilitated design process is proposed, and it ensures the consistent failure analysis among the product design process.

3- A real-life product is studied.